\documentclass[twocolumn,showpacs,preprintnumbers,prb,amsmath,amssymb]{revtex4}
\usepackage{graphicx}
\usepackage{dcolumn}
\usepackage{bm}
\usepackage{longtable}
\usepackage{hyperref}

\begin{document}

\title{Spin Fine Structure in Optically Excited Quantum Dot Molecules}

\author
{M. Scheibner$^{1}$} \email{scheibner@bloch.nrl.navy.mil}
\author{M. F. Doty$^{1}$}
\author{I. V. Ponomarev$^{1}$}
\author{A.S. Bracker$^{1}$}
\author{E.A. Stinaff$^{1}$}
\author{V.L. Korenev$^{2}$}
\author{ T.L. Reinecke$^{1}$}
\author{D. Gammon$^{1}$}
\affiliation{$^{1}$Naval Research
Laboratory, Washington, DC 20375, USA} \affiliation{$^{2}$A.F.
Ioffe Physical Technical Institute, St. Petersburg 194021 Russia}


\date{\today}
\begin{abstract}
The interaction between spins in coupled quantum dots is revealed
in distinct fine structure patterns in the measured optical
spectra of InAs/GaAs double quantum dot molecules containing zero,
one, or two excess holes. The fine structure is explained well in
terms of a uniquely molecular interplay of spin exchange
interactions, Pauli exclusion and orbital tunneling. This
knowledge is critical for converting quantum dot molecule
tunneling into a means of optically coupling not just orbitals but
also spins.
\end{abstract}

\maketitle

\section{Introduction}
\begin{widetext}

\begin{figure}[h!]
\centering
\includegraphics[width=16.2cm]{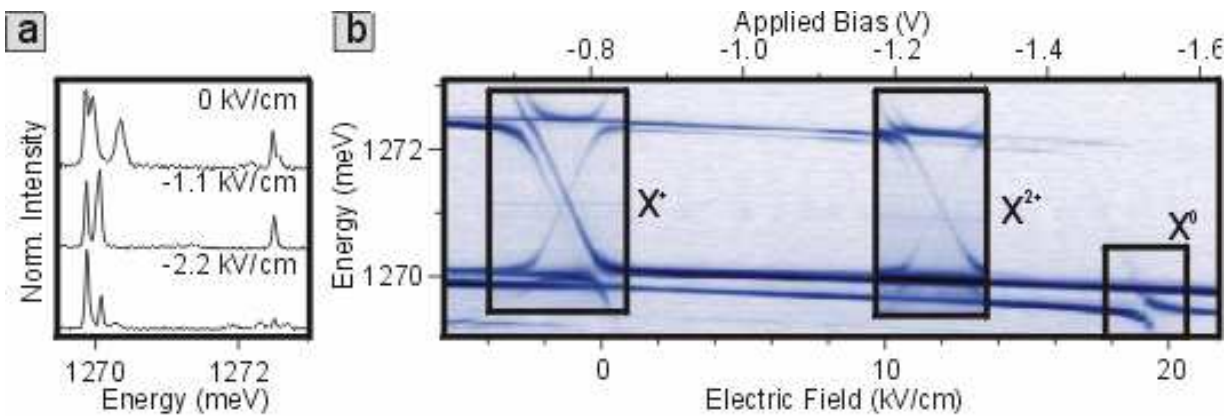}
\caption{\label{fig:Overview}(Color online) \textbf{(a)}
Conventional PL line spectra at selected electric fields.
\textbf{(b)} Fine structure of molecular resonances in the PL
transitions of the neutral exciton ($X^0$), positive trion ($X^+$)
and doubly positively charged exciton ($X^{2+}$) in a QDM is
revealed clearly by a 2-dimensional intensity color scale plot of
the PL signal. Here the bottom dot has a vertical height of
$h_B=4$ nm and the top dot has a vertical height of $h_T=2.5$ nm
with a dot
separation of $d=4$ nm.} 
\end{figure}
\end{widetext}

Exchange coupling between spins in a double quantum dot molecule
(QDM) is an essential component for spin-based quantum information
\cite{Loss98, DiVincenzoNature00, Aws02}. Rapid progress has been
made recently in doubly charged QDMs that are measured and
controlled electrically \cite{PettaScience05, KoppensScience05}. A
corresponding understanding of the spin-spin interactions in
optically controlled QDMs is not yet available. These systems
could lead to ultrafast, wireless control of spin qubits and
optical entanglement of two spins in two dots. To this end it is
now critical to obtain a detailed measurement and understanding of
the spin states of optically excited QDMs.
\\
\indent In optically excited quantum dots (QDs), an electron-hole
(\textit{e-h}) pair is created in the presence of the previously
existing spin(s). As we will show, the resulting \textit{e-e},
\textit{h-h}, and \textit{e-h} exchange interactions determine the
spin states and can be directly measured through fine structure in
the spectra \cite{GammonPRL96, Urbaszek03}. Moreover, electron or
hole levels may be coupled by carrier tunneling in a QDM, with the
orbital wavefunctions continuously tuneable from atomic to
molecular in nature. Recently, photoluminescence (PL) spectra of
vertically-stacked InAs/GaAs QDMs measured as a function of
electric field have led to the clear identification of tunnel
coupling in neutral\cite{Krenner05, OrtnerPRL05, StinaffScience06,
NakaokaPRB06, DotyPRL06, BrackerAPL06} and charged
QDMs.\cite{StinaffScience06, Krenner06, PonomarevPSSB06,
DotyPRL06} In these earlier studies some limited information on
the spin fine structure can be found. Here we give the broad
picture of the spin fine structure physics in QDMs
\\
\indent Of great importance is the case of a doubly charged QDM
because of its potential use as a two qubit system. To understand
this system, we examine cases of zero, one and two charges in the
QDM. We demonstrate that all of the observed fine structure
features of these systems arise from combinations of three
fundamental quantum mechanical processes -- tunneling, exchange,
and Pauli exclusion. An ideal tool to visualize this underlying
physics are PL spectra as function of electric field (see Fig.
\ref{fig:Overview}). Such two dimensional color scale plots reveal
clearly a
\clearpage \noindent rich variety of fine structure in the PL
transitions of the different excitonic charge states. The primary
goal of this paper is to explain the origin of the various fine
structure patterns observed at the anticrossing points in such
spectra.
\\
\section{Synopsis}
\begin{figure}[b]
\centering
\includegraphics[width=8.6cm]{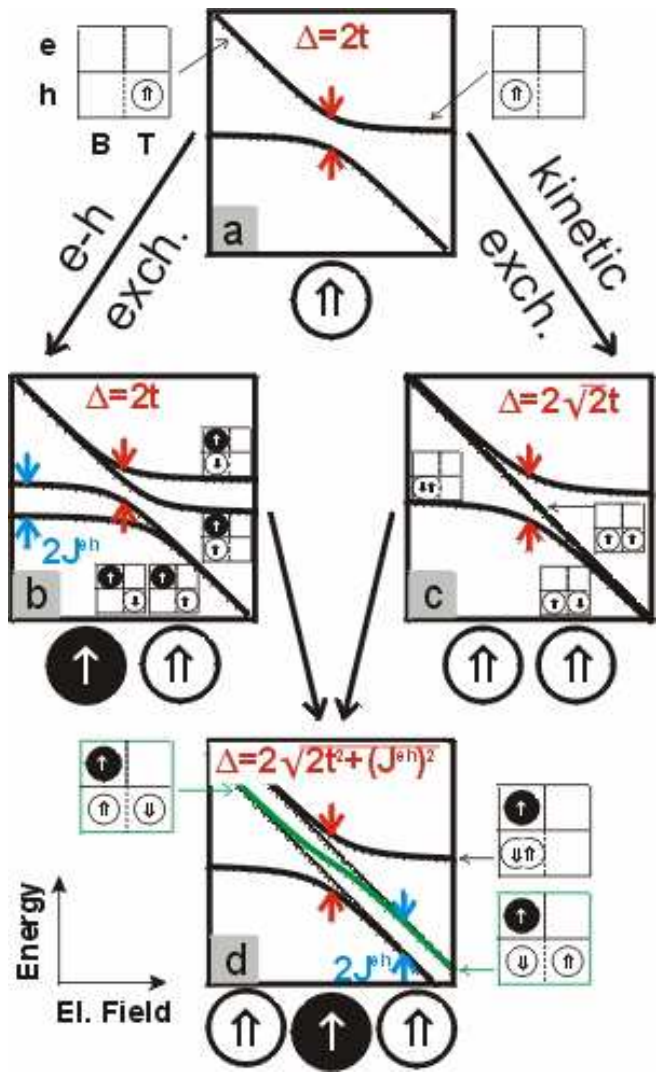}
\caption{\label{fig:Summary}(Color online) Schematic energy
dispersions diagrams (state energy vs. electric field) for a QD
molecular resonance of spin carrying particles in two spatial
configurations of: \textbf{(a)} A single (spin carrying) particle,
\textbf{(b)} Two spin carrying particles of different type (e.g.,
electron and hole), where only one of them tunnels, \textbf{(c)}
Two holes, where both may share the same orbital state, and
\textbf{(d)} Three spin carrying particles, e.g. two holes and one
electron which does not take part in the tunneling process. In the
$\boxplus$ the two left squares represent the bottom QD's and the
two right squares the top QD's valence and conduction band. The
$\bullet$'s and $\circ$'s denote electrons and holes respectively.
}
\end{figure}

\indent Here we study spin-spin interactions of (pseudo-)spin-1/2
particles, namely electrons and holes, in the vicinity of
molecular resonances in double-dot QDMs. Four fundamental cases
can be identified. These are depicted in Fig. \ref{fig:Summary}.
\\
\indent The first and simplest case (Fig. \ref{fig:Summary}(a)) is
the molecular resonance of a single spin-carrying particle -- in
this example, a hole. In the absence of a magnetic field, all
states are two-fold degenerate. The hole can occupy either of the
two QDs, as indicated in the $\boxplus$ diagrams. The potential
energies of the two configurations change relative to each other
with electric field, because the two QDs are located at a slightly
different position within an applied electric field. Without loss
of generality we chose the energy of the configuration with the
hole in the bottom QD to be field independent. As the energies of
the two configurations approach each other, the hole wavefunction
begins to form bonding (lower energy) and anti-bonding (higher
energy) orbital states between the two QDs. In the field dependent
energy diagram, Fig. \ref{fig:Summary}(a), this results in an
avoided crossing of the energy levels of the two charge
configurations. The energy splitting between the bonding and
antibonding states is twice the tunnel coupling constant
($\Delta=2t$). In the experiment this case is realized by a QDM
which contains a single hole. It is observed as the final state in
the optical recombination of a positively charged exciton (trion).
\\
\indent The behavior which results from adding another spin
depends on the nature of the spin carrying particle. If an
electron and a hole occupy the QDM, the physics of the fine
structure splitting across a molecular resonance of one of the two
particles is as depicted in Fig. \ref{fig:Summary}(b).
\textit{e-h} exchange interaction ($J^{eh}$) leads to a small
splitting between the states with parallel and antiparallel spins.
With both spins in the same QD the interaction between the spins
is maximum. The splitting reduces across the molecular resonance
as one spin tunnels into the second QD. For both spin
configurations (parallel and antiparallel) the splitting between
bonding and anti-bonding states is twice the tunneling constant
($\Delta=2t$). Examples for this type of behavior are the neutral
exciton and the doubly charged exciton. There the spin-spin
interaction is characterized by the \textit{e-h} exchange energy
$J^{eh}$.
\\
\indent Figure \ref{fig:Summary}(c) captures the physics in the
third case, where both particles are of the same type, e.g. two
holes, and may share the same orbital state. At a molecular
resonance the Pauli principle demands that tunneling into the same
QD can only occur if the particles form a spin-singlet. The three
(degenerate) spin-triplet states are unaffected by a molecular
resonance. Consequently, tunneling splits singlet from triplet
states (kinetic exchange). Note, only one of three degenerate
triplet states is depicted in Fig. \ref{fig:Summary}(c). The
anticrossing splitting of the singlet states amounts to
$\Delta=2\sqrt{2}t$. This is larger by a factor of $\sqrt{2}$ than
for a single particle and results from the indistinguishability of
the two particles. Physical implementations of this case are, for
example, QDMs occupied by two holes, or two electrons, or a
neutral biexciton with either the two electrons or the two holes
spin paired and restricted to one of the QDs.
\\
\indent The fourth case combines the previous two cases (Fig.
\ref{fig:Summary}(d)). It is obtained by considering, for example,
two holes and one electron, where the holes may tunnel but the
location of the electron is fixed to the bottom QD. Here the
\textit{e-h} exchange interaction lifts the degeneracy of the
three hole spin triplet states in the region of the molecular
resonance. While the two triplet states with parallel spins remain
unaffected by the resonance (dashed lines), the third triplet
state couples now to the singlet states and `wiggles' through the
resonance region (green line). The total anticrossing splitting in
this case amounts to $\Delta=2\sqrt{2t^2+2(J^{eh})^2}$. By
following the green line in Fig. \ref{fig:Summary}(d) through the
resonance it is found that the two holes swap their spin. This
exemplifies that the interplay of tunneling and spin interactions
can in principle be used to manipulate spin states. The singly
charged excitons (trions) are examples where such a situation is
realized experimentally.
\\
\indent In the following we will give the details of the
experimental observation of these spin fine structure patterns. In
particular we will examine the neutral exciton, $X^0$, and the
doubly positively charged exciton, $X^{2+}$, which illustrate
\textit{e-h} exchange in the presence of hole tunneling. The
optical transition of the $X^{2+}$ leaves the QDM in a 2-hole
state and thus also shows the kinetic \textit{h-h} exchange
splitting, which is created by tunneling and Pauli exclusion. For
the singly charged exciton, $X^+$, we find that the interplay of
all these processes results indeed in the `wiggling' of a spectral
line, which is a signature of the mixing of spin singlet and spin
triplet states. Note, although all experiments are performed with
hole tunneling, we find in other experiments that electron
tunneling yields qualitatively the same physics.
\\
\section{Samples and Methods}
\begin{figure}[b]
\centering
\includegraphics[width=8.8cm]{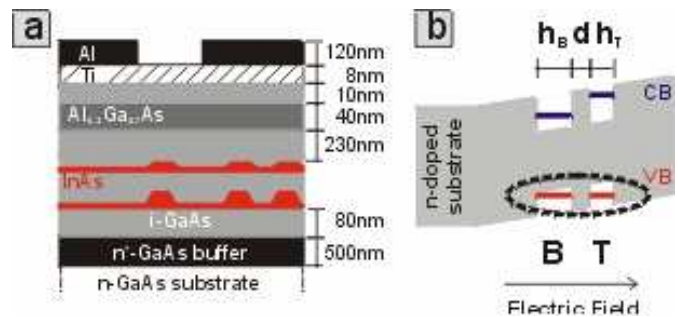}
\caption{\label{fig:Sample}(Color online) \textbf{(a)} Schematic
of the sample layer sequence. The n$^+$-GaAs buffer layer was
doped with Tellurium at $\approx 10^{15}$ cm$^{-3}$. \textbf{(b)}
Schematic of the QDM diode level structure.}
\end{figure}
\indent We form QDMs by the subsequent growth of two closely
spaced layers of self assembled InAs/GaAs QDs. Because of strain
the QDs of the second layer nucleate preferentially on top of QDs
in the first layer. The QDMs are embedded in the insulating region
of an n-I Schottky diode grown by molecular beam epitaxy on top of
a (100) n-type GaAs substrate wafer. The layer sequence is
depicted in Fig. \ref{fig:Sample}(a). In order to produce a
gradient of the QD density across the wafer the rotation of the
sample during growth was halted for the formation of the QD
layers. This ensures that we obtain the right density for studying
individual molecules in each sample growth. The electric contacts
on top of the samples were formed by a semi transparent titanium
layer and an aluminum mask with 1 $\mu$m size apertures for the
optical access. We control the heights of the QDs by an indium
flush technique. The self-assembled QDs are partially capped with
GaAs followed by an increase in temperature in order to
redistribute and partly remove the still exposed part of the QD
material. For the QDMs discussed here the nominal heights of the
bottom dot, $h_B$, and the top dot, $h_T$, were chosen so that
predominantly the bottom dot exhibits the lower transition energy,
i.e., $h_T\leq h_B$ \cite{BrackerAPL06}. This allows the hole
levels to be brought into resonance with a positive electric
field, while the electron level of the top dot ($T$) is shifted to
much higher energies relative to that of the bottom dot ($B$) (see
Fig. \ref{fig:Sample}(b)). In this case, we include in our
discussion only the ``atomic'' s-shell orbital states for the
electron localized in the bottom dot and for the holes in both
dots. We consider samples with relatively wide interdot barriers
($d\geq4$ nm) such that hole tunneling rates are small ($\leq 1$
meV).\cite{BrackerAPL06}
\\
\indent The case of electron tunneling is realized using the same
principle. It is achieved by either a reversed dot
order\cite{Krenner05} or by applying a negative electric
field.\cite{BrackerAPL06} In the latter case a p-type instead of
an n-type substrate is required for the proper pinning of the
Fermi level in the Schottky diode.
\\
\indent Apertures in the aluminum mask with a diameter of about 1
$\mu$m allow us to address and detect individual QDMs optically.
The QDMs were excited below the energy of the wetting layer with a
frequency tunable continuous wave titanium sapphire laser. A
liquid nitrogen cooled charge coupled device (CCD) camera in
combination with a 1200 mm$^{-1}$ line grating inside a 0.55 m
monochromator provided an overall spectral resolution of 50
$\mu$eV for the detection of the PL. The samples were measured at
a temperature of 10 K. The interplay of optical excitation,
recombination, and tunneling to and from the substrate leads to
the observation of several charge states in the same PL
spectrum.\cite{WarbourtonNature00, Ware05}
\\
\section{Nomenclature}
\indent To describe the quantum states of a QDM we use
$^{e_{B}e_{T}}_{h_{B}h_{T}}X^{q}$, where $e_{B(T)}$ and $h_{B(T)}$
are the number of electrons and holes in the respective dot and
$q$ is the total charge. For example, $^{10}_{21}X^{2+}$ for a
doubly positively charged exciton, i.e., 1 electron and 2 holes in
the bottom dot and 0 electron and 1 hole in the top dot. Likewise
$^{\underline{1}0}_{2\underline{1}}X^{2+}$ corresponds to the
interdot transition, $^{10}_{21}X^{2+}\Rightarrow$
$^{00}_{20}h^{2+}$, where $h^{2+}$ is a 2-hole state. We label
specific spin states of a charge configuration as, e.g.,
$^{\hspace{3pt}\uparrow\hspace{3pt},\hspace{0.5pt}0}_{\Uparrow\Downarrow,\Uparrow}X^{2+}$,
where we use the fact that a hole in the ground state of a QD can
take only two spin projections ($\Uparrow,\Downarrow\hspace{5pt}
\equiv\pm3/2$), similar to the case of the spin-1/2 electron
($\uparrow,\downarrow$) \cite{KavokinPRB04}. We specify the hole
spin singlet and triplet configurations as, e.g.,
$^{\uparrow\hspace{0.5pt},\hspace{0.5pt}0}_{\Uparrow,\Downarrow}X^+_S$
and
$^{\uparrow\hspace{1pt},\hspace{0.5pt}0}_{\Uparrow,\Downarrow}X^+_T$
\cite{Z-SingletTripletNotation}. These various state
configurations correspond to an ``atomic''-like basis in which the
spatial wavefunctions are localized predominantly in one or the
other QD. They are realized away from the molecular resonances. In
the resonance regions the ``atomic'' states are mixed by tunneling
into ``molecular'' states. The electrons and holes are treated as
non-identical particles with an explicit exchange interaction
between them.
\\
\section{Theoretical Description}
\indent The few-particle Hamiltonians that describe these
excitonic states consist of three parts. The first is a sum of
electron and hole single-particle QD Hamiltonians
$\widehat{\mathbf{h}}^{e(h)}$, the second describes Coulomb
interactions, and the third is a short-range electron-hole
(\textit{e-h}) exchange: $ A\sum_{i,j}\delta({\bf r}_{ei}-{\bf
r}_{hj})\hat{\sigma}^e_{zi}\hat{\sigma}^h_{zj}$, where $A$ is the
exchange amplitude. We use three s-shell orbitals: the bottom dot
electron and hole states and the top dot hole or electron state.
The fourth s-shell orbital is tuned far out of resonance and can
therefore be neglected.\\
\indent The Coulomb terms are:
\begin{equation}
\label{eqn:Hgen} V^{\alpha,\beta}_{ijkl}=\pm\int d{\bf r} d{\bf
r}'|{\bf r}-{\bf r}'|^{-1} \varphi^{\alpha*}_i({\bf
r})\varphi^{\beta*}_k({\bf r}') \varphi^{\alpha}_j({\bf
r})\varphi^{\beta}_l({\bf r}'),
\end{equation}
\newline
with the orthonormalized wavefunction $\varphi^{\alpha}_{i}$ of
particle $\alpha$ in dot $i$. In the wide barrier limit only the
direct Coulomb interactions, e.g. the repulsion of two holes in
the bottom dot $V^{hh}_{BB} \equiv V^{hh}_{BBBB}$, or between the
two dots, $V^{hh}_{BT} \equiv V^{hh}_{BBTT}$ are non-negligible.
In addition only \textit{e-h} exchange within the bottom dot with
$J^{eh} \equiv
J^{eh}_{BB}=A\int\mathit{d}\mathbf{r}\left|\varphi_B^e\right|^{2}\left|\varphi_B^{h}\right|^2$
has to be considered. A theoretical discussion of the general
case, including the narrow barrier limit, will be given
elsewhere.\cite{Ponomarev_unpub}
\\
\indent In the following the bottom QD is chosen as an energy
reference point. This means, that the energies of configurations
with no charges in the top QD are independent of the applied
electric field, $F$, while those with one hole (electron) in the
top dot change with field proportional to $f\equiv
pF=|e|(d+(h_B+h_T)/2)F$, and those with 2 holes (electrons) change
proportional to $2f$. Here $e$ is the elementary charge, $d$ is
the separation of the dots and $h_B$, $h_T$ are the heights of the
dots. Thus, $p$ is equivalent to the dipole moment of two
elementary charges separated by $d$. It follows that intradot
transitions, which keep the number of holes (electrons) in the top
QD constant, have field-independent PL lines, while interdot
transitions have slope $\pm p$.
\\
\indent Nine parameters are physically relevant -- the six Coulomb
interactions $V^{hh}_{BB}$, $V^{eh}_{BB}$, $V^{ee}_{BB},
V^{hh}_{BT}$ ($V^{ee}_{BT}$), $V^{eh}_{BT}$ and $V^{hh}_{TT}$
($V^{ee}_{TT}$), the intradot \textit{e-h} exchange interaction,
$J^{eh}_{BB}$, the tunneling matrix element $t_{h(e)}=-\langle
\varphi^{h(e)}_{B}|\widehat{\mathbf{h}}^{h(e)}|\varphi^{h(e)}_{T}\rangle$,
and the dipole moment $p$. Maximally four parameters are required
in order to fit the spectral patterns, which are obtained by
dispersing the optical transition spectrum of a given excitonic
charge state in electric field.  These parameters are an effective
Coulomb potential, which is given by a combination of the above
Coulomb terms, the \textit{e-h} exchange, the tunneling rate and
the dipole moment.
\\
\indent In the course of this paper we will give the Hamiltonians
of the respective charge states in matrix representation. For
example, a single hole in a QDM represented by the two basis
states $\left|1\right>=$ $^{00}_{10}h^+$ and $\left|2\right>=$
$^{00}_{01}h^+$ is described by:
\begin{equation}\label{eqn:singlehole}
H^h\equiv\left(%
\begin{array}{cc}
  H_{11} & H_{12} \\
  H_{21} & H_{22} \\
\end{array}%
\right)=\left(%
\begin{array}{cc}
  0 & t_h \\
  t_h & -f \\
\end{array}%
\right).
\end{equation}
For the multi particle cases we will choose mainly basis states,
$\left|i\right>$, with diagonal matrix elements, $H_{ii}$, that
contain the field dependence, \textit{e-h} exchange, and some
combination of the above Coulomb potentials specific to the
respective charge configuration. These can be associated easily
with the (atomic) energy levels away from the molecular
resonances. In this case coupling between the basis states
$\left|i\right>$ and $\left|j\right>$ will be represented by
nonzero off-diagonal matrix elements ($H_{ij}$ with $i\neq j$)
proportional to $t_{h(e)}$. The off-diagonal ``tunneling'' terms
in the Hamiltonian lead to the formation of molecular states.
\\
\indent This theoretical description yields qualitatively the same
results for the two cases of electron and hole tunneling. Both
cases differ only in the magnitude of the respective matrix
elements and the sign of the field dependence. In our experiments
we have so far not found any indication for a qualitative
difference between electron and hole tunneling. Because the
tunneling rates are larger\cite{BrackerAPL06} and PL transitions
overlap more strongly in the case of electron tunneling, we will
focus in the following on the case of hole level resonances in
order to give a clear picture of the underlying physics.
\\
\section{Electron-Hole Exchange}
\subsection{The neutral exciton}
\begin{figure}[t]
\centering
\includegraphics[width=8.6cm]{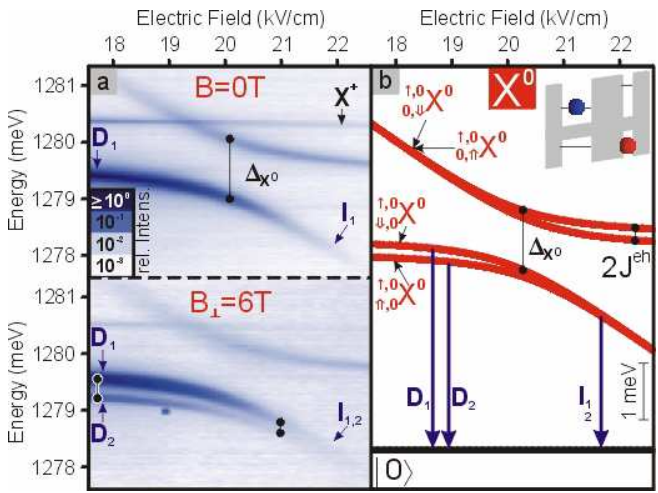}
\caption{\label{fig:Exciton}(Color online) Exciton ($X^0$) in an
uncharged QDM ($h_B=h_T=2.5$ nm, $d=4$ nm). \textbf{(a)} The PL
spectrum as a function of electric field -- at zero magnetic field
($B=0$) (top) and in a transverse magnetic field ($B=6$ T)
(bottom) \textbf{(b)} The calculated energy diagram
($\Delta_{X^0}=2t_{h}=1.05$ meV, $2J^{eh}=240$ $\mu$eV). The
direct exciton and its corresponding transition ($D_{1,2}\equiv$
$^{\underline{1}0}_{\underline{1}0}X^0$) has the hole in the same
dot as the electron and its energy is almost independent of
electric field, while the indirect exciton ($I_{1,2}\equiv$
$^{\underline{1}0}_{0\underline{1}}X^0$) has the hole in the
opposite dot and its energy changes linearly with field. An
anticrossing occurs when their energies are resonant.}
\end{figure}
\indent We consider first the case of \textit{e-h} exchange for
the neutral exciton, $X^0$ $\equiv$ 1 electron + 1 hole, in a QDM
(Fig. \ref{fig:Exciton}). The exciton has spin states that are
optically allowed (bright, $\uparrow\Downarrow$ and
$\downarrow\Uparrow$), or optically forbidden (dark,
$\uparrow\Uparrow$ and
$\downarrow\Downarrow$).\cite{TischlerPRBr02, BayPRB02} In what
follows we will not list spin degenerate states in which all
spins, including the electron spin, are flipped. At zero magnetic
field a single intradot exciton line,
$^{\underline{1}0}_{\underline{1}0}X^0$, anticrosses with the
relatively weak interdot transition,
$^{\underline{1}0}_{0\underline{1}}X^0$, with a splitting of
$\Delta_{X^0}=1.0$ meV (Fig. \ref{fig:Exciton}(a) top). When a
transverse magnetic field of $B=6$ T is applied (Voigt geometry) a
second, normally dark, intradot spectral line appears 320 $\mu$eV
lower in energy (Fig.\ref{fig:Exciton}(a) bottom), because the
transverse magnetic field mixes the bright and dark states
\cite{BayPRB00}. The bright-dark splitting of the intradot
exciton, $^{10}_{10}X^0$, arises from e-h exchange, similar to the
case of a single dot. As the intradot exciton evolves into the
interdot exciton through the anticrossing region, the \textit{e-h}
exchange splitting decreases substantially because of the
decreased overlap of the electron and hole wavefunctions in the
interdot configuration, $^{10}_{01}X^0$.
\\
\indent This physics is captured by calculations of the energy
level structure as seen in Fig. \ref{fig:Exciton}(b), where we
have included \textit{e-h} exchange, $J^{eh}$, only for the
intradot configurations $^{10}_{10}X^0$. Here the Hamiltonian of
the bright ($\left|1\right>=$ $^{\uparrow0}_{\Downarrow0}X^0$,
$\left|2\right>=$ $^{\uparrow0}_{0\Downarrow}X^0$) and dark
($\left|3\right>=$ $^{\uparrow0}_{\Uparrow0}X^0$,
$\left|4\right>=$ $^{\uparrow0}_{0\Uparrow}X^0$) excitons at zero
magnetic field are
\begin{equation}
\label{eqn:X0} \widehat{H}^{X^0}=\left(%
\begin{array}{cccc}
   J^{eh} & t_{h} & 0 & 0 \\
   t_{h} & -f & 0 & 0 \\
   0 & 0 & -J^{eh} & t_{h} \\
   0 & 0 & t_{h} & -f\\
\end{array}%
\right),
\end{equation}
where energy and field are relative to the center of the exciton
anticrossing and $t_{h}$ is the single hole tunneling rate
\cite{Z-tunnelingrate}. The applied electric field, $F$, changes
the hole energy in the top QD relative to that in the bottom QD by
$-f$. The neutral exciton exemplifies that in a QDM \textit{e-h}
exchange is large within the same dot but small between dots.
\\
\subsection{The doubly charged exciton}
\indent When we have two or more electrons (or holes) we must also
consider \textit{e-e} (or \textit{h-h}) interactions. The case of
two electrons has been discussed for dots controlled by electrical
gating. \cite{BurkhardPRB00, PettaScience05, KoppensScience05} The
case of two holes is qualitatively the same and is observed
optically in the doubly positively charged exciton ($X^{2+}$). In
contrast to the uncharged exciton, the doubly charged exciton
transition shows an `x'-pattern \cite{StinaffScience06} with four
anticrossings that involve two direct (\textit{A} and \textit{D})
and two indirect (\textit{B} and \textit{C}) transitions, as seen
in Fig. \ref{fig:Xpp}(a). This transition pattern is understood
using the calculated energy diagrams for the initial $X^{2+}$
states and the final 2-hole states left after recombination, which
both have hole level resonances (Fig.\ref{fig:Xpp}(b)).
\\
\indent Each anticrossing has fine structure whose pattern
provides a signature of the corresponding spin configurations. The
two anticrossings on the left in the spectrum of Fig.
\ref{fig:Xpp}(a) arise from the doubly charged exciton states
($\Delta_{X^{2+}}$) and those on the right from the 2-hole states
($\Delta_{h^{2+}}$). We first consider the initial states of the
$X^{2+}$ transitions, as seen at the top of Fig. \ref{fig:Xpp}(b).
With respect to spin, the $X^{2+}$ states are qualitatively the
same as the states of the neutral exciton. In particular,
\textit{e-h} exchange is present only when the electron shares the
bottom dot with the unpaired hole, $^{10}_{12}X^{2+}$. The
\textit{e-h} exchange splitting decreases as the bottom QD is
filled with two spin paired holes (total spin zero) and the top QD
is left with an unpaired hole, $^{10}_{21}X^{2+}$. Note that this
is visible in the $X^{2+}$ transitions even without magnetic field
(see Fig. \ref{fig:Xpp}(a) top left corner).
\\
\indent This is described by the Hamiltonian for the basis states
$\left|1\right>=$
$^{\hspace{2.5pt}\uparrow\hspace{3pt},\hspace{0.5pt}0}_{\Downarrow\Uparrow,\Downarrow}X^{2+}$,
$\left|2\right>=$
$^{\hspace{0.5pt}\uparrow\hspace{0.5pt},\hspace{3pt}0}_{\Downarrow,\Uparrow\Downarrow}X^{2+}$
and $\left|3\right>=$
$^{\hspace{2.5pt}\uparrow\hspace{3pt},\hspace{0.5pt}0}_{\Uparrow\Downarrow,\Uparrow}X^{2+}$,
$\left|4\right>=$
$^{\hspace{0.5pt}\uparrow\hspace{0.5pt},\hspace{3pt}0}_{\Uparrow,\Downarrow\Uparrow}X^{2+}$,
\begin{equation}
\label{eqn:Xpp} \widehat{H}^{X^{2+}}=\left(%
\begin{array}{cccc}
  -f & t_{h} & 0 & 0 \\
  t_{h} & E-2f+J^{eh} & 0 & 0 \\
  0 & 0 & -f & t_{h} \\
  0 & 0 & t_{h} & E-2f-J^{eh} \\
\end{array}%
\right),
\end{equation}
\newline
where $E=V^{eh}_{BB}-V^{eh}_{BT}+V^{hh}_{TT}-V^{hh}_{BB}$. Note,
that this is similar to the Hamiltonian of the neutral exciton.
\\
\section{`Kinetic' Exchange: two-hole states}
\begin{figure}[t!]
\centering
\includegraphics[width=8.0cm]{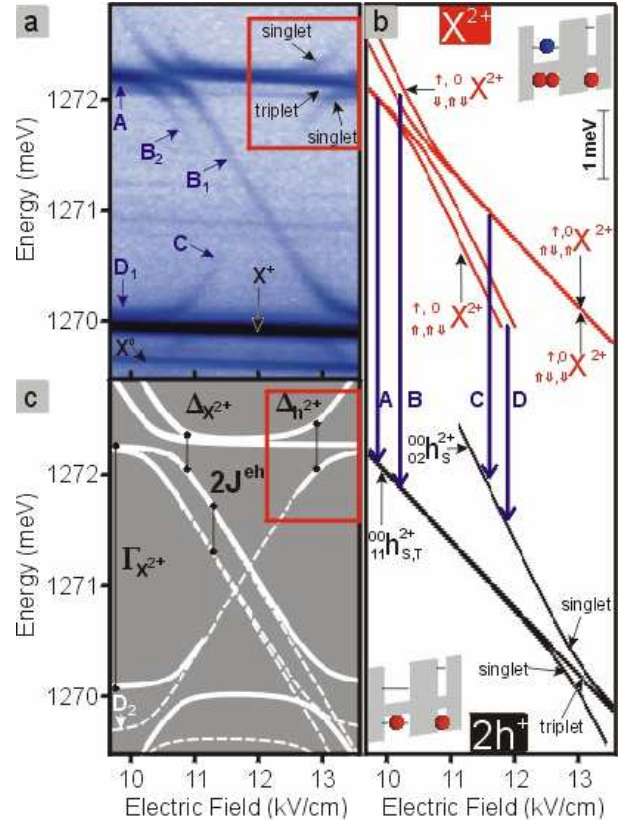}
\caption{\label{fig:Xpp}(Color online) Doubly positively charged
exciton ($X^{2+}$) in a QDM ($h_B=4$ nm, $h_T=2.5$ nm, $d=4$ nm).
\textbf{(a)} In the measured PL spectrum as a function of electric
field an `x'-shaped pattern is formed by the $X^{2+}$ transitions
($A\equiv$ $^{\underline{1}0}_{\underline{2}1}X^{2+}$, $B\equiv$
$^{\underline{1}0}_{1\underline{2}}X^{2+}$, $C\equiv$
$^{10}_{21}X^{2+}\rightarrow$ $^{00}_{02}h^{2+}$, $D\equiv$
$^{\underline{1}0}_{\underline{1}1}X^{2+}$). \textbf{(b)} The
calculated level diagram contains the states of the $X^{2+}$ (red
lines) and the 2-hole states ($h^{2+}$) (black lines). The
calculation is done using a fit to four parameters -- $t_h$,
$J^{eh}$, $\Gamma_{X^{2+}}$\cite{Z-Gamma} and $p$ 
($\Delta_{h^{2+}}=2\sqrt{2}t_{h^{+}}=410$ $\mu$eV,
$\Delta_{X^{2+}}=2t_{h^{+}}=310$ $\mu$eV, $2J^{eh}=410$ $\mu$eV,
$\Gamma_{X^{2+}}=2.19$ meV, $p=0.99$ meV/(kV/cm)). Note that
another 2-hole state in which both holes are in the bottom dot
exists, but is not seen in the displayed energy range. The two
different field dependences seen in the 2-hole and $X^{2+}$ states
lead to the field-independent direct (\textit{A} and \textit{D})
and field-dependent indirect (\textit{B} and \textit{C})
transitions. \textit{B} and \textit{D} are split by the
\textit{e-h} exchange $J^{eh}$ of the optically excited state
($X^{2+}$). The red box in (a) marks an area where the
singlet-triplet splitting of the resonance of the 2-hole states in
(b) is reproduced nicely by the PL signal. The anticrossing
appears rotated because it occurs in the final state of the
transitions (\textit{A} and \textit{C}). \textbf{(c)} Calculated
PL spectrum. The solid lines resemble the spectrum in (a). The
dashed lines map transitions that are optically weak or forbidden
by optical selection rules.}
\end{figure}
\indent If for the final state both holes are in the same dot
($^{00}_{02}h^{2+}$) the Pauli principle requires that there can
be only a spin singlet state. On the other hand, if the holes are
each in a different dot ($^{00}_{11}h^{2+}$) there will be a
singlet and three triplet states.\cite{Z-2holebotdot} For large
barrier width ($d\geq4$ nm) the interdot \textit{h-h} exchange is
negligible, and the spin singlet and triplet states are degenerate
at electric fields away from the anticrossing region. However,
because tunneling conserves spin, only the interdot singlet
configuration
$^{\hspace{2pt}0\hspace{2pt},\hspace{0.5pt}0}_{\hspace{1.5pt}\Uparrow\hspace{1.5pt},\Downarrow}h^{2+}_S$
mixes with the intradot singlet
$^{0,\hspace{3pt}0}_{0,\Uparrow\Downarrow}h^{2+}_S$, and the
degenerate triplet states $^{00}_{11}h^{2+}_T$ pass through
unaffected as shown in Fig \ref{fig:Xpp}(b). 
Mixing of the singlet states in the anticrossing region leads to
an effective or ``kinetic'' \cite{Fazekas99} \textit{h-h} exchange
splitting between the singlet and the three degenerate triplets,
even though interdot \textit{h-h} exchange is negligible for wide
barriers. This type of exchange, which is highly sensitive to the
applied electric field through the resonant tunneling, provides a
basis for externally manipulating the spin
coupling.\cite{PettaScience05}
\\
\indent The full Hamiltonian of two holes in a QDM is given in
appendix \ref{sec:appendix1}. The reduced Hamiltonian for the two
singlet states, $\left|1\right>=$ $^{00}_{02}h^{2+}_S$ and
$\left|2\right>=$ $^{00}_{11}h^{2+}_S$, that are relevant here is
\begin{equation}
\label{eqn:hhs} \widehat{H}^{h^{2+}}_S=\left(\begin{array}{cc}
  V^{hh}_{TT}-2f & \sqrt{2}t_{h} \\
  \sqrt{2}t_{h} & V^{hh}_{BT}-f \\
\end{array}\right),
\end{equation}
The factor of $\sqrt{2}$ in the tunneling rate between the singlet
states comes from the fact that two indistinguishable holes can
tunnel.\cite{StinaffScience06} The three triplet configurations,
$^{00}_{11}h^{2+}_T$, have the same energy as the singlet basis
state $^{00}_{11}h^{2+}_S$, namely $V^{hh}_{BT}-f$. They are fully
decoupled.
\\
\indent Thus, the kinetic \textit{h-h} exchange splitting arises
in the 2-hole energy diagram, and the \textit{e-h} exchange
splitting arises in the $X^{2+}$ energy diagram. Therefore, the
transition spectrum, which is the difference between the two
energy diagrams, shows separately both types of spin fine
structure (Fig. \ref{fig:Xpp}(a) and (c)). On the right are the
2-hole spin patterns, and on the left are the the $X^{2+}$ spin
patterns.
\\
\indent Note, some transitions are optically weak (e.g.
$B_2\equiv$
$^{\hspace{0.5pt}\underline{\uparrow}\hspace{0.5pt},\hspace{3pt}0}_{\Uparrow,\underline{\Downarrow}\Uparrow}X^{2+}$)
or forbidden by the optical selection rules  (e.g. $D_2\equiv$
$^{\hspace{0.5pt}\underline{\uparrow}\hspace{0.5pt},\hspace{3pt}0}_{\underline{\Uparrow},\Downarrow\Uparrow}X^{2+}$)
(white dashed lines in Fig. \ref{fig:Xpp}(c)), but become visible
in the vicinity of anticrossings. There they gain some oscillator
strength from states with optically stronger transitions.
\\
\section{Interplay of electron-hole and kinetic Exchange: singly charged exciton}
\begin{figure}[t!]
\centering
\includegraphics[width=8.0cm]{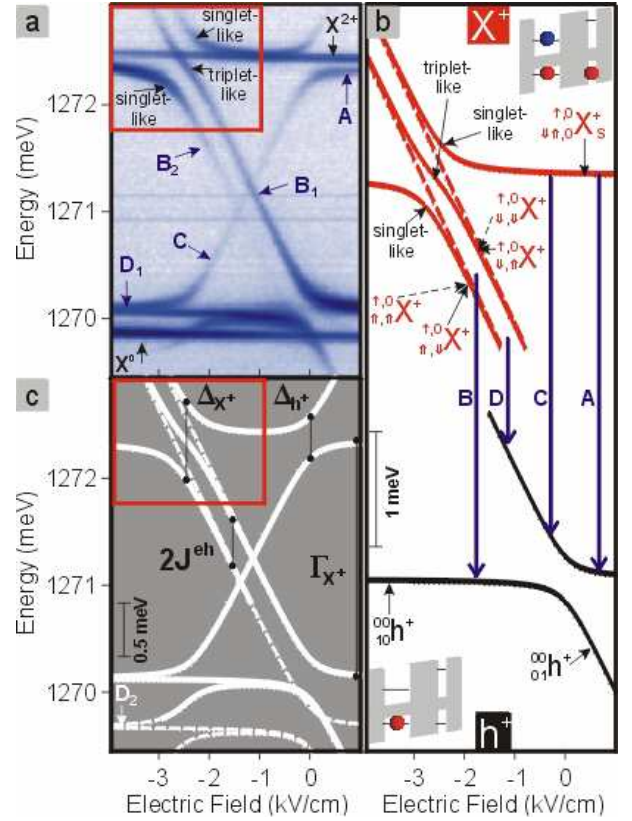}
\caption{\label{fig:Xp}(Color online) \textbf{(a)} In the measured
PL spectrum as a function of electric field an `x'-shaped pattern
is formed by the $X^+$ transitions ($A\equiv$
$^{\underline{1}0}_{\underline{2}0}X^+$, $B\equiv$
$^{\underline{1}0}_{1\underline{1}}X^+$, $C\equiv$
$^{10}_{11}X^+\rightarrow$ $^{00}_{01}h^+$, $D\equiv$
$^{\underline{1}0}_{\underline{1}1}X^+$). \textbf{(b)} The
calculated level diagram contains the states of the positive trion
($X^+$) (red lines) and the states of a single hole ($h^+$) (black
lines) ($\Delta_{h^+}=2t_{h^+}=390$ $\mu$eV,
$\Delta_{X^+}=2\sqrt{2t_{h}^2+(J^{eh})^2}=750$ $\mu$eV,
$2J^{eh}=430$ $\mu$eV, $\Gamma_{X^+}=2.21$ meV,\cite{Z-Gamma}
$p=0.99$ meV/(kV/cm)). Note that another $X^+$ state in which both
holes are in the top dot exists, but is not seen in the displayed
energy range. The two different field dependences seen in the
$h^+$ and $X^{+}$ states lead to the field-independent direct
(\textit{A} and \textit{D}) and field-dependent indirect
(\textit{B} and \textit{C}) transitions. \textit{B} and \textit{D}
are split by the \textit{e-h} exchange $J^{eh}$ of the optically
excited state ($X^{+}$). The red box in (a) marks an area where
the singlet-triplet mixing at the $X^+$ resonance in (b) is
reproduced nicely by the PL signal. In the lower left corner of
(a) the anticrossing is reproduced by the transitions \textit{D}
and \textit{C} with different intensity ratios, but is partially
covered by $^{\underline{1}0}_{\underline{1}0}X^0$. \textbf{(c)}
Calculated PL spectrum. The solid lines resemble the spectrum in
(a). The dashed lines map transitions that are optically weak or
forbidden by optical selection rules. }
\end{figure}

\indent With the $X^{+}$ transitions we probe the QDM when it is
charged with a single hole (see Fig. \ref{fig:Xp}). In the $X^+$
states kinetic \textit{h-h} exchange and \textit{e-h} exchange are
both present and compete to determine the character of the spin
state. In Fig. \ref{fig:Xp}(a) the spectral pattern for the
positive trion, $X^+$, is shown. This pattern can be readily
understood using the energy state diagrams in Fig. \ref{fig:Xp}(b)
of both the trion and the hole that is left behind after
recombination. The discovery of this overall ``x''-pattern and its
identification was made in Ref. 10. However, the spin fine
structure was only partially interpreted. Now we are able to
complete the description of the measured fine structure -- at
least at the level of the current experimental resolution. We
focus our discussion on the anticrossing pattern in the box in
Fig. \ref{fig:Xp}(a), in which an apparent triplet transition
wiggles as it passes through the resonance.
\\
\indent At electric fields away from the anticrossing region,
intradot \textit{e-h} exchange determines the spin structure of
the $X^+$. That is, as shown in the top of Fig. \ref{fig:Xp}(b),
\textit{e-h} exchange leads to a fine structure doublet with a
splitting of $2J^{eh}$, much like the intradot $X^0$ case. The
higher energy component consists of the electron and one hole in
the bottom dot with their spins antiparallel
($^{\hspace{0.5pt}\uparrow\hspace{0.5pt},\hspace{0.5pt}0}_{\Downarrow,\Uparrow}X^+$,
$^{\hspace{0.5pt}\uparrow\hspace{0.5pt},\hspace{0.5pt}0}_{\Downarrow,\Downarrow}X^+$),
while the lower energy component consists of parallel electron and
hole spin in the bottom dot
($^{\hspace{0.5pt}\uparrow\hspace{0.5pt},\hspace{0.5pt}0}_{\Uparrow,\Downarrow}X^+$,
$^{\hspace{0.5pt}\uparrow\hspace{0.5pt},\hspace{0.5pt}0}_{\Uparrow,\Uparrow}X^+$).
\cite{Z-SingletTripletNotation, Z-OtherBasis_prb}
\indent As the electric field is tuned through the $X^+$
anticrossing region, tunnel coupling with the singlet
$^{\hspace{3pt}\uparrow\hspace{3pt},0}_{\Uparrow\Downarrow,0}X^+_S$
state forces the spin states
($^{\hspace{0.5pt}\uparrow\hspace{0.5pt},\hspace{0.5pt}0}_{\Downarrow,\Uparrow}X^+$,
$^{\hspace{0.5pt}\uparrow\hspace{0.5pt},\hspace{0.5pt}0}_{\Uparrow,\Downarrow}X^+$)
to form a hole spin singlet-like state
($^{\hspace{0.5pt}\uparrow\hspace{0.5pt},\hspace{0.5pt}0}_{\Downarrow,\Uparrow}X^+_S$)
and a hole spin triplet-like state
($^{\hspace{0.5pt}\uparrow\hspace{0.5pt},\hspace{0.5pt}0}_{\Downarrow,\Uparrow}X^+_T$).
This triplet would pass straight through the resonance (as with
the 2-hole states) except that \textit{e-h} exchange continues to
couple it to the singlets, causing it to shift continuously
between the asymptotes determined by the \textit{e-h} exchange
splitting outside the anticrossing region. Essentially, in passing
through the anticrossing region (from right to left) the
$^{\hspace{0.5pt}\uparrow\hspace{0.5pt},\hspace{0.5pt}0}_{\Downarrow,\Uparrow}X^+$
state evolves continuously into the
$^{\hspace{0.5pt}\uparrow\hspace{0.5pt},\hspace{0.5pt}0}_{\Uparrow,\Downarrow}X^+$
state through this triplet-like state.
\\
\indent The Hamiltonian that describes this behavior of the basis
states $\left|1\right>=$
$^{\hspace{3pt}\uparrow\hspace{3pt},0}_{\Uparrow\Downarrow,0}X^+_S$,
$\left|2\right>=$
$^{\hspace{0.5pt}\uparrow\hspace{0.5pt},\hspace{0.5pt}0}_{\Downarrow,\Uparrow}X^+$
and $\left|3\right>=$
$^{\hspace{0.5pt}\uparrow\hspace{0.5pt},\hspace{0.5pt}0}_{\Uparrow,\Downarrow}X^+$
is \cite{Z-OtherBasis_prb}
\begin{equation}
\label{eqn:Xp}
\widehat{H}^{X^+}_{\Uparrow\Downarrow}=\left(%
\begin{array}{ccc}
  E_{B} & t_h & t_h \\
  t_h & E_{BT}-f+J^{eh} & 0 \\
  t_h & 0 & E_{BT}-f-J^{eh} \\
\end{array}%
\right).
\end{equation}
The complete Hamiltonian for all six basis states, i.e. also
including $^{\uparrow,\hspace{3pt}0}_{0,\Uparrow\Downarrow}X^+_S$,
$^{\hspace{0.5pt}\uparrow\hspace{0.5pt},\hspace{0.5pt}0}_{\Downarrow,\Downarrow}X^+$
and
$^{\hspace{0.5pt}\uparrow\hspace{0.5pt},\hspace{0.5pt}0}_{\Uparrow,\Uparrow}X^+$,
is given in appendix \ref{sec:appendix1}. With energy and field
measured relative to the anticrossing of the hole states, $E_{B}$
is the energy of
$^{\hspace{3pt}\uparrow\hspace{3pt},0}_{\Uparrow\Downarrow,0}X^+_S$
and 
$E_{BT}=E_{X_B}-V^{eh}_{BT}-V^{eh}_{BB}+V^{hh}_{BB}-V^{hh}_{BT}$. 
Again not all transitions between
the $X^{+}$ states and the $h^{+}$ states are observed in the
measured PL spectrum, because they are optically weak (e.g.
$^{\hspace{0.5pt}\underline{\uparrow}\hspace{0.5pt},\hspace{0.5pt}0}_{\Downarrow,\underline{\Downarrow}}X^+$
($B_1$)) or forbidden (e.g. $D_2\equiv$
$^{\hspace{0.5pt}\underline{\uparrow}\hspace{0.5pt},\hspace{0.5pt}0}_{\underline{\Uparrow},\Uparrow}X^+$)
(Fig. \ref{fig:Xp}(c), white dashed lines).
\\
\indent An alternative representation of the $X^+$ as compared to
Eq. (\ref{eqn:Xp}) can be given in terms of the basis
$\left|1\right>=$ $^{\hspace{3pt} \uparrow
\hspace{2.5pt},0}_{\Uparrow \Downarrow,0}X^+_S$, $\left|2\right>=$
$^{\uparrow \hspace{0.5pt}, \hspace{0.5pt}0}_{\Uparrow,
\Downarrow}X^{+}_{S}$ and $\left|3\right>=$ $^{\uparrow
\hspace{0.5pt}, \hspace{0.5pt}0}_{\Uparrow,
\Downarrow}X^{+}_{T}$\cite{StinaffScience06}. The transformation
of Eq. (\ref{eqn:Xp}) to this basis yields
\begin{equation}
\label{eqn:Xpalt} \widehat{H}^{X^+}_{\Uparrow\Downarrow}=\left(
\begin{array}{ccc}
  E_{B} & \sqrt{2}t_h & 0 \\
  \sqrt{2}t_h & E_{BT}-f & J^{eh} \\
  0 & J^{eh} & E_{BT}-f \\
\end{array}
\right).
\end{equation}
In this representation it becomes obvious that the \textit{e-h}
exchange couples the triplet $^{\uparrow \hspace{0.5pt},
\hspace{0.5pt}0}_{\Uparrow, \Downarrow}X^{+}_{T}$ to the singlets.
\\
\indent Tunneling and \textit{e-h} exchange lead to a measured
anticrossing energy $\Delta_{X^+}=2\sqrt{2t_{h}^2+(J^{eh})^2}$.
The remaining hole spin triplet states
($^{\hspace{0.5pt}\uparrow\hspace{0.5pt},\hspace{0.5pt}0}_{\Downarrow,\Downarrow}X^+$,
$^{\hspace{0.5pt}\uparrow\hspace{0.5pt},\hspace{0.5pt}0}_{\Uparrow,\Uparrow}X^+$)
retain their character and pass unaffected through the coupling
region as shown in Fig. \ref{fig:Xp}(b) (red dashed lines),
because Pauli blocking prevents the holes from tunneling. Thus, at
the anticrossing point there is a kinetic exchange splitting
between singlet- and triplet-like states, but \textit{e-h}
exchange splits the degeneracy between the triplet states and
leads to a mixing between the singlets and one of the triplets.
\\
\indent Interestingly, the singlet-triplet mixing observed here is
similar to that found in transport studies.\cite{PettaScience05,
KoppensScience05} In that case hyperfine interactions, which are
different in the two QDs, break the singlet-triplet symmetry of
two electrons away from resonance. Here it is the \textit{e-h}
exchange with the localized electron in the bottom QD, that breaks
the singlet-triplet symmetry away from resonance. In both cases
the strong tunneling term partially restores the singlet-triplet
symmetry. However a residual mixing remains that results in the
triplet ``wiggling''.
\\
\section{Conclusions}
\indent We have shown that the fine structure patterns measured in
the optical spectra of QDMs are understood in detail in terms of
the interplay between spin exchange, Pauli principle and tunneling
in the limit of wide barriers and negligible direct interdot
exchange. Our description applies equally well to electron
tunneling and negatively charged QDMs. We note that we have also
measured the neutral biexciton, which is found to have spin
structure qualitatively similar to the $X^{2+}$ as
expected.\cite{ScheibnerXX07} Interesting but more subtle effects
such as fine structure due to asymmetries (e.g. lateral
displacement of the dots) have for the most part remained below
the resolution of our measurements. In cases where the barrier is
relatively thin\cite{StinaffScience06, DotyPRL06, BrackerAPL06}
such that wave function overlap becomes large, we expect
additional interactions (such as direct interdot exchange) to
become significant.
\\
\indent An important implication of these results is that exchange
is effectively turned off (or on) when the QDM is optically
excited to specific spin states, thereby providing the opportunity
for an ultrafast single qubit or 2-qubit operation.  For example,
the kinetic exchange interaction that splits the triplet and
singlet states of the 2-holes, $^{00}_{11}h^{2+}$, could be
`optically gated' for a well defined time by driving the QDM up
and down through a $^{10}_{21}X^{2+}$ state (i.e. a virtual
$^{\underline{1}0}_{\underline{2}1}X^{2+}$ transition).
\\
\appendix
\section{Electric field values of molecular resonances}
\begin{figure}[t]
\centering
\includegraphics[width=8cm]{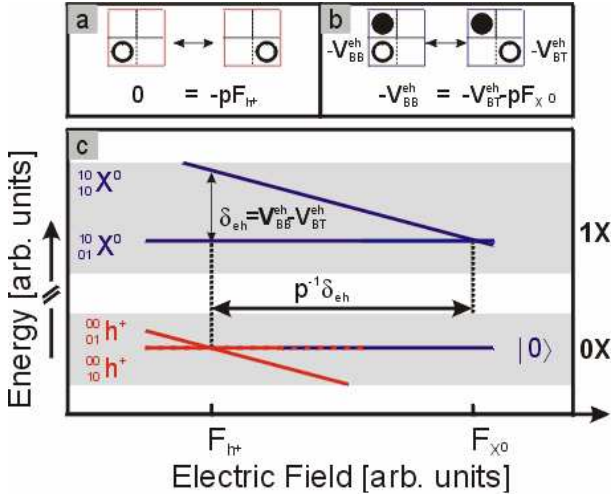}
\caption{\label{fig:Coulomb-shift}(Color online)(\textbf{a})
Condition for a hole level resonance if the QDM is occupied by a
hole only. (\textbf{b}) Condition for a hole level resonance if
the QDM is occupied by a hole and an electron that is fixed to the
bottom QD. (\textbf{c}) Geometrical interpretation of the
separation between both resonances.}
\end{figure}

\indent Anticrossings are observed in the field dependent spectral
map of a QDM when interdot transitions cross intradot transitions
for a given charge state (see Fig. \ref{fig:Overview}).
\cite{StinaffScience06, Krenner05} At the anticrossing, a hole
(electron) becomes delocalized across both dots, forming bonding
and antibonding molecular states. Different charge configurations
become energetically degenerate (or resonant) at different applied
electric fields because of different Coulomb interactions.
\\
\indent The relative position in electric field where the
corresponding anticrossings occur in the spectrum depend on the
Coulomb interactions between the involved charges. Figure
\ref{fig:Coulomb-shift} demonstrates this schematically for the
cases when the QDM system is occupied by one hole only and when it
is occupied by one hole and one electron. If we tune the ground
state hole levels of the bottom and the top QD in resonance, the
hole can be in either one of the dots while the electron remains
localized in the bottom dot only. With only one hole no Coulomb
interactions have to be considered and the electric field needed
to tune the hole levels into resonance has to account only for the
offset between the hole levels at zero applied field. Without loss
of generality we set the field at which this resonance occurs to
zero (Fig. \ref{fig:Coulomb-shift}(a)). With the additional
electron in the bottom QD we have to account for the \textit{e-h}
Coulomb interaction in both, intradot and interdot, charge
configurations -- $V^{eh}_{BB}$ and $V^{eh}_{BT}$. At resonance
the hole has the same potential energy in both configurations
(here we ignore the formation of bonding and antibonding states),
and it follows that the corresponding electric field is
proportional to the difference of the two Coulomb terms.
\begin{equation}
\label{eqn:FX0} F_{X^0}=(V^{eh}_{BB}-V^{eh}_{BT})/p=\delta_{eh}/p
\end{equation}
$\delta_{eh}$ is the energy difference between the two charge
configurations at the resonance of the bare hole (see Fig.
\ref{fig:Coulomb-shift}(c)). Equivalently it is the energy which
is required to move one hole from the bottom to the top QD while
an electron resides in the bottom QD. \\
\indent In Eq. (\ref{eqn:FX0}) we have neglected \textit{e-h}
exchange, $J^{eh}$. If we take it into account Eq. (\ref{eqn:FX0})
reads:
\begin{equation}
\label{eqn:deltaeh} F_{X^0}=(\delta_{eh}\pm J^{eh})/p.
\end{equation}
\\
The relative electric field of any other level resonance can be
found in analogy. For example, for the hole level resonance of the
$X^{2+}$ we obtain\cite{ScheibnerICPS06}
\begin{equation}
\label{eqn:deltaehXpp}
F_{X^{2+}}=(\delta_{eh}+V^{hh}_{TT}-V^{hh}_{BB}\pm J^{eh})/p.
\end{equation}
The fact that in Fig. \ref{fig:Overview} the molecular resonance
of the $X^{2+}$ occurs at an electric field between the molecular
resonances of the bare hole and the neutral exciton indicates that
in this molecule $V^{hh}_{TT}$ is smaller than $V^{hh}_{BB}$.
\\
\section{6$\times$6 Hamiltonians of the 2-holes \& the positive trion}
\label{sec:appendix1} \indent All three singlet and three triplet
states of the two holes are described by a 6$\times$6 Hamiltonian.
In the basis $\left|1\right>=$
$^{0,\hspace{3pt}0}_{0,\Uparrow\Downarrow}h^{2+}_S$,
$\left|2\right>=$
$^{\hspace{0.5pt}0\hspace{0.5pt},\hspace{0.5pt}0}_{\Uparrow,\Downarrow}h^{2+}_S$,
$\left|3\right>=$
$^{\hspace{3pt}0\hspace{2.5pt},0}_{\Uparrow\Downarrow,0}h^{2+}_S$,
$\left|4\right>=$
$^{\hspace{0.5pt}0\hspace{0.5pt},\hspace{0.5pt}0}_{\Uparrow,\Uparrow}h^{2+}_T$,
$\left|5\right>=$
$^{\hspace{0.5pt}0\hspace{0.5pt},\hspace{0.5pt}0}_{\Uparrow,\Downarrow}h^{2+}_T$
and $\left|6\right>=$
$^{\hspace{0.5pt}0\hspace{0.5pt},\hspace{0.5pt}0}_{\Downarrow,\Downarrow}h^{2+}_T$
it is
\begin{widetext}
\begin{equation}
\label{eqn:hhsfull}
\widehat{H}^{h^{2+}}=\left(\begin{array}{cccccc}
  V^{hh}_{TT}-2f & \sqrt{2}t_{h} & 0 & 0 & 0 & 0 \\
  \sqrt{2}t_{h} & V^{hh}_{BT}-f & \sqrt{2}t_{h} & 0 & 0 & 0 \\
  0 & \sqrt{2}t_{h} & V^{hh}_{BB} & 0 & 0 & 0 \\
  0 & 0 & 0 & V^{hh}_{BT}-f & 0 & 0 \\
  0 & 0 & 0 & 0 & V^{hh}_{BT}-f & 0 \\
  0 & 0 & 0 & 0 & 0 & V^{hh}_{BT}-f \\
\end{array}\right),
\end{equation}
In Eq. (\ref{eqn:Xp}) we suppressed the basis states
$^{\uparrow,\hspace{3pt}0}_{0,\Uparrow\Downarrow}X^+_S$,
$^{\hspace{0.5pt}\uparrow\hspace{0.5pt},\hspace{0.5pt}0}_{\Downarrow,\Downarrow}X^+$
and
$^{\hspace{0.5pt}\uparrow\hspace{0.5pt},\hspace{0.5pt}0}_{\Uparrow,\Uparrow}X^+$.
The 6$\times$6 Hamiltonian for the $X^+$ in the basis
$\left|1\right>=$
$^{\uparrow,\hspace{3pt}0}_{0,\Uparrow\Downarrow}X^+_S$,
$\left|2\right>=$
$^{\hspace{3pt}\uparrow\hspace{3pt},0}_{\Uparrow\Downarrow,0}X^+_S$,
$\left|3\right>=$
$^{\hspace{0.5pt}\uparrow\hspace{0.5pt},\hspace{0.5pt}0}_{\Downarrow,\Uparrow}X^+$,
$\left|4\right>=$
$^{\hspace{0.5pt}\uparrow\hspace{0.5pt},\hspace{0.5pt}0}_{\Uparrow,\Downarrow}X^+$,
$\left|5\right>=$
$^{\hspace{0.5pt}\uparrow\hspace{0.5pt},\hspace{0.5pt}0}_{\Downarrow,\Downarrow}X^+$
and $\left|6\right>=$
$^{\hspace{0.5pt}\uparrow\hspace{0.5pt},\hspace{0.5pt}0}_{\Uparrow,\Uparrow}X^+$
is:
\begin{equation}
\label{eqn:Xpfull}
\widehat{H}^{X^+}_{\Uparrow\Downarrow}=\left(%
\begin{array}{cccccc}
  E_T-2f & 0 & t_h & t_h & 0 & 0\\
 0 & E_{B} & t_h & t_h & 0 & 0\\
  t_h & t_h & E_{BT}-f+J^{eh} & 0 & 0 & 0\\
  t_h & t_h & 0 & E_{BT}-f-J^{eh} & 0 & 0 \\
  0 & 0 & 0 & 0 & E_{BT}-f+J^{eh} & 0 \\
  0 & 0 & 0 & 0 & 0 & E_{BT}-f-J^{eh} \\
\end{array}%
\right),
\end{equation}
where
$E_T=E_B+2(V^{eh}_{BB}-V^{eh}_{BT})+V^{hh}_{TT}-V^{hh}_{BB}$. In
terms of the alternative basis, $\left|1\right>=$
$^{\uparrow,\hspace{3pt}0}_{0,\Uparrow\Downarrow}X^+_S$,
$\left|2\right>=$ $^{\hspace{3pt} \uparrow
\hspace{2.5pt},0}_{\Uparrow \Downarrow,0}X^+_S$, $\left|3\right>=$
$^{\uparrow \hspace{0.5pt}, \hspace{0.5pt}0}_{\Uparrow,
\Downarrow}X^{+}_{S}$, $\left|4\right>=$ $^{\uparrow
\hspace{0.5pt}, \hspace{0.5pt}0}_{\Uparrow, \Downarrow}X^{+}_{T}$,
$\left|5\right>=$
$^{\hspace{0.5pt}\uparrow\hspace{0.5pt},\hspace{0.5pt}0}_{\Downarrow,\Downarrow}X^+$
and $\left|6\right>=$
$^{\hspace{0.5pt}\uparrow\hspace{0.5pt},\hspace{0.5pt}0}_{\Uparrow,\Uparrow}X^+$
the Hamiltonian of the $X^+$ reads:
\begin{equation}
\label{eqn:Xpaltfull}
\widehat{H}^{X^+}_{\Uparrow\Downarrow}=\left(
\begin{array}{cccccc}
  E_T-2f & 0 & \sqrt{2}t_h & 0 & 0 & 0\\
  0 & E_{B} & \sqrt{2}t_h & 0 & 0 & 0 \\
  \sqrt{2}t_h & \sqrt{2}t_h & E_{BT}-f & J^{eh} & 0 & 0 \\
  0 & 0 & J^{eh} & E_{BT}-f & 0 & 0 \\
  0 & 0 & 0 & 0 & E_{BT}-f+J^{eh} & 0 \\
  0 & 0 & 0 & 0 & 0 & E_{BT}-f-J^{eh} \\
\end{array}
\right).
\end{equation}
\end{widetext}

\begin{acknowledgments}
We would like to acknowledge the financial support by NSA/ARO,
CRDF, RFBR, RSSF, and ONR. M.F.D., I.V.P and E.A.S, are NRC/NRL
Research Associates.
\end{acknowledgments}

\bibliography{SpinXQDM-full-AL}

\begin{thebibliography}{31}
\bibitem[{Z-t()}]{Z-tunnelingrate}
\bibinfo{note}{For wide barriers contributions to the tunneling rate due to the
  presence of additional charges are negligible \cite{StinaffScience06}.}
\bibitem[{Z-2()}]{Z-2holebotdot}
\bibinfo{note}{{The $^{00}_{20}h^{2+}$ state is at higher energy and is ignored
  here.}}
\bibitem[{Z-O()}]{Z-OtherBasis_prb}
\bibinfo{note}{{We define: $^{\hspace{0.5pt} \uparrow \hspace{0.5pt},
  \hspace{0.5pt}0}_{\Downarrow, \Uparrow}X^{+} = \frac{1}{\sqrt{2}}
  \left(^{\uparrow \hspace{0.5pt}, \hspace{0.5pt}0}_{\Uparrow,
  \Downarrow}X^{+}_{S} + {^{\uparrow \hspace{0.5pt},\hspace{0.5pt}0}_{\Uparrow,
  \Downarrow}X}^{+}_{T} \right)$ and $^{\hspace{0.5pt} \uparrow \hspace{0.5pt},
  \hspace{0.5pt}0}_{\Uparrow, \Downarrow}X^{+} = \frac{1}{\sqrt{2}}
  \left(^{\uparrow \hspace{0.5pt}, \hspace{0.5pt}0}_{\Uparrow,
  \Downarrow}X^{+}_{S} - {^{\uparrow \hspace{0.5pt},\hspace{0.5pt}0}_{\Uparrow,
  \Downarrow}X}^{+}_{T} \right)$ }}.
\end{thebibliography}

\clearpage

\end{document}